\documentstyle[12pt,cite,axodraw,epsfig,epsf]{article} 
\newcommand{\be}{\begin{equation}}
\newcommand{\ee}{\end{equation}}
\newcommand{\etal}{{\em et al.}}
\newcommand{\eg}{{\em e.g.}}

\newcommand{\eqref}[1]{Eq.~(\ref{#1})}

\begin{document}

%----------------Convenient shorthands for LaTeX commands-----------%
\def\be{\begin{equation}}
\def\bea{\begin{eqnarray}}
\def\ee{\end{equation}}
\def\eea{\end{eqnarray}}
                              \def\barr{\begin{array}}
                              \def\earr{\end{array}}
\def\dis{\displaystyle}
%----------------------------Text macros----------------------------%
\def\eg{{\em e.g.}}
\def\etc{{\em etc.}}
\def\etal{{et al.}}

\def\ie{{\em i.e.}}
\def\viz{{\em viz.}}
%---------------------------Math macros-----------------------------%
\def\lsim{\:\raisebox{-0.5ex}{$\stackrel{\textstyle<}{\sim}$}\:}
\def\gsim{\:\raisebox{-0.5ex}{$\stackrel{\textstyle>}{\sim}$}\:}
                              \def\mev{\: \rm MeV} 
                              \def\gev{\: \rm GeV} 
                              \def\tev{\: \rm TeV} 
                              \def\pb {\: \rm pb}
                              \def\fb {\: \rm fb}
\def\gappeq{\mathrel{\rlap {\raise.5ex\hbox{$>$}}
            {\lower.5ex\hbox{$\sim$}}}}
\def\lappeq{\mathrel{\rlap{\raise.5ex\hbox{$<$}}
            {\lower.5ex\hbox{$\sim$}}}}
\def\ra{\rightarrow}
\def\mand{\qquad {\rm and} \qquad}
                              \def\ptsl{p_T \hspace{-1.1em}/\;}
                              \def\pslash{p \hspace{-0.6em}/\;}
                              \def\msl{m \hspace{-0.8em}/\;}
\def\rp{R\!\!\!/ _p}
\def\nut{\chi^0}
\def\cha{\chi^\pm}
\def\squ{\tilde q}
\def\slep{\tilde l}
\def\slashiii#1{\setbox0=\hbox{$#1$}#1\hskip-\wd0\hbox to\wd0{\hss\sl/\/\hss}}
\def\slashiv#1{#1\llap{\sl/}}
\newcommand{\rpv}{$\slashiii{R}$}
%===================================================================%
\begin{flushright}
{\large \tt MRI-P-010301}\\
{\large \tt hep-ph/0103122}
\end{flushright}

\vspace*{2ex}

\begin{center}
{\Large\bf Heavy quark production via supersymmetric interaction at
 a neutrino factory}

\vskip 15pt

{\sf Debrupa Chakraverty\footnote{rupa@mri.ernet.in}, 
Anindya Datta\footnote{anindya@mri.ernet.in} and 
Biswarup Mukhopadhyaya\footnote{biswarup@mri.ernet.in}}

\vskip 5pt
{\sf Harish Chandra Research Institute, 
Chhatnag Road, Jhusi, Allahabad 211 019, India.} 

\vskip 15pt
{\Large\bf Abstract}
\end{center}

We investigate b-quark production in both charged and neutral current
channels through $\nu_\mu-N$ scattering at a neutrino factory,
mediated by the lepton flavour violating interactions present in a
supersymmetric theory with broken R-parity. Using values of the
effective interaction strengths well below the current and projected
experimental bounds, we are still able to predict markedly enhanced
event rates, especially for the neutral current events which are not
allowed at the lowest order in the standard model (SM). Data from
neutrino factories can therefore be used to probe strengths of such
interactions to considerably higher precision than what can be
envisioned in other experiments.


\vskip 20pt
%\hspace*{0.65cm}

%\def\baselinestretch{1.0}

%===================================================================%

%\newpage

\renewcommand{\thefootnote}{\arabic{footnote}}
\setcounter{footnote}{0}

It is being frequently suggested nowadays that a neutrino factory,
cashing on the intense and well-calibrated supply of $\nu_\mu$'s and
$\nu_e$'s coming out of a muon storage ring
\cite{storage_ring,fat_rev}, can go a long way in investigating the
world of neutrinos where numerous puzzles are still in store for us.
In addition to its usefulness in probing neutrino oscillations, such
high precision neutrino experiments can have several other interesting
physics goals \cite{fat_rev,physics_goal}. One of these is the possible
investigation of physics beyond the standard model, something which
becomes a necessity once one accepts the existence of neutrino mass
and mixing. It is thus natural to ask whether there are observables,
rising above the threshold of detectability in a neutrino factory,
which will unequivocally imply the existence of such new physics
interactions involving the neutrino sector.  In this paper we suggest
heavy flavour production, particularly in neutral current events, as
such a demonstration of new physics.  We illustrate our point in the
context of an R-parity violating supersymmetric theory, emphasizing,
however, that the conclusions drawn therefrom can be of a rather
general nature.

We assume a design for muon production, capture, cooling, acceleration
and storage as given in ref. \cite{storage_ring,fat_rev}.  The muons
decaying along a straight section of the storage ring will give rise
to a collimated neutrino beam that is of interest to us. It has been
argued that one can thus have an yearly supply of a few times
$10^{20}$ neutrinos (or antineutrinos) of either flavour.  The
feasibility studies carried out so far agree that the muons in the
storage ring can easily have energies upto 50 GeV.  In order to
observe new physics effects in the deep inelastic scattering (DIS) of
these neutrinos, it is preferable to have a near-site detector rather
than a long-baseline one, so that oscillation effects do not dominate.

The interaction taking place at the detector (which is the fixed
target for the scattering phenomena under study) is basically between
the neutrinos and the partons present in nucleons.  In the Standard
Model (SM), heavy quark production in such scattering can take place
at the tree-level via the charged current channels, $\nu_\mu {\bar u}
\to \mu^- {\bar b}$, $\nu_\mu {\bar c} \to \mu^- {\bar b}$, $\nu_\mu d
\to \mu^- c$ and $\nu_\mu s \to \mu^- c$. The SM charged current
cross-section depends on two factors, the Cabibbo-Kobayashi-Maskawa
(CKM) matrix elements and the quark distributions in a nucleon.  All
the processes mentioned above, excepting the third one, occur through
the interaction of muon-neutrino with the sea quarks in nucleons and
are thus depend on the parton distributions relevant in the energy
range under consideration. As far as charm production is concerned,
the rates undergo suppression by either the strange quark distribution
or the CKM element $V_{cd}$. For b-production, however, there is no
CKM-diagonal channel, and the decidedly small element $V_{ub}$
\cite{pdg} severely suppresses the predicted rates. In addition,
neutral current processes (whose measurements are also projected goals
of a neutrino factory) where a b-quark can be tagged among the DIS
products can come only at the one-loop level at the SM, and therefore
such processes are unlikely to be detected, given any realistic muon
luminosity.
 
 One can conclude from above that if there is any observation of an
 excess in charged current b-production over the standard model rate,
 or if neutral current events are observed with  a b in the final
 state, it will clearly signal some kind of new physics. The
 question is, given other kinds of constraints (especially those from
 b-decays) on the same operators that give rise to such events, is it
 possible at a neutrino factory to have any observable excess of
 events, or to use the absence of such excess to impose useful bounds
 on the new interactions?

 The conservation of both baryon and lepton number in the SM is a
 consequence of its gauge current structure and renormalizability.  In
 the supersymmetric (SUSY) extension of the SM \cite{susy}, the
 existence of scalar quarks and leptons make it possible to violate
 baryon and lepton numbers, without leading to any theoretical
 problems.  This entails the possibility of the violating R-parity
 in SUSY \cite{rpv}, defined by $R=(-1)^{3 B + L + 2 S}$, where $B$ is
 the baryon number, $L$, the lepton number and $S$, the spin of
 the particle.  Clearly, all SM particles have $R = +1$ while their
 superpartners have $R = -1$. While there is no fundamental principle
 dictating the conservation of R-parity, the need to avoid fast proton
 decay leads to the conjecture that out of $B$ and $L$, only {\em one}
 can be violated. Here we consider the situation where $L$-violation
 occurs, since that is the scenario which can affect the observed
 events at a neutrino factory. The resulting interactions can in
 general fake the signals of neutrino oscillation, as
 discussed in an earlier work \cite{rp_fake}.
 
 In terms of the superfields, the $\Delta L = 1$ part of the $R$-violating 
superpotential is given by
\cite{rpv_pheno},
\be
{\cal L}_{R\!\!\!/}=\epsilon_{ab}\left[\epsilon_i {\hat L}_i^a {\hat H}_u^b +
\lambda_{ijk} ({\hat L}_i^a {\hat L}_j^b) {\hat E}_k^c +
\lambda_{ijk}' ({\hat L}_i^a {\hat Q}_j^b) {\hat D}_k^c \right]
\end{equation} 
where ${\hat L}_i$ and ${\hat Q}_i$ are $SU(2)$ 
doublet leptons and quarks,
 ${ \hat E}^c_i$ and ${ \hat D}^c_i$ are $SU(2)$ singlet charged 
lepton and down quarks, and 
${\hat H}_u$ is the Higgs superfield responsible for the 
generation of the up-type quark masses.  $i$, $j$ and $k$ 
are generation indices, while colour indices are 
not explicitly shown above. The corresponding interaction terms 
can then be written in terms of the component 
fields as \cite{rpv_pheno}
\bea
{\cal L}_{R\!\!\!/} &=&  \lambda'_{ijk} ~\big[ ~\tilde d^j _L \,\bar d ^k _R \nu^i _L
  + (\tilde d ^k_R)^\ast ( \bar \nu ^i_L)^c d^j _L +
   \tilde \nu ^i _L \bar d^k _R d ^j _L  \nonumber \\
& & ~~~~~~~ -\tilde e^i _L \bar d ^k _R u^j _L
- \tilde u^j _L \,\bar d ^k _R e^i _L
-(\tilde d^k _R)^\ast (\bar e ^i _L)^c u^j _L \big] + h.c. \nonumber \\
& & + \lambda_{ijk} ~\big[ ~\tilde e^j _L \,\bar e ^k _R \nu^i _L
  + (\tilde e ^k_R)^\ast (\bar \nu ^i _L)^c e^j _L +
   \tilde \nu ^i _L \bar e^k _R e ^j _L  - (i \leftrightarrow j) \big] + h.c
\label{eq:rp}
\eea                                          
 
The most important thing to note among the terms shown above is that
not only lepton number but also {\em lepton flavour} can be violated
through them. Thus it is possible to have $\Delta L = 0$ processes
with flavour violation by taking suitable products of the $\Delta L =
1$ couplings.  One of the consequences of this is the production of the 
b-quark through both charged and flavour changing neutral currents
(FCNC), represented by diagrams shown in figure 1. Although 9
$\lambda$-type and 27 $\lambda^{'}$-type couplings are allowed
altogether, we demonstrate our argument by assuming non-zero values of
only the $\lambda^{'}$-interactions with indices conforming to the
requirements of b-production in $\nu_\mu$ scattering.

 As has been already mentioned, we shall confine ourselves to
 near-site experiments in which one places the neutrino detectors at a
 short distance ($40~m$) from the straight section of the storage ring
 \cite{fat_rev}. Under such circumstances, the incoherent scattering
 effects (only of $\nu_\mu$'s here, by virtue of our choice of
 non-vanishing new physics interactions) dominate over oscillation
 effects. The number of b-production events per year, via either
 charged-current or neutral current interactions, can be obtained by
 folding the relevant cross-section by a survival probability and the 
 neutrino flux,

 \be 
 N_Q={\frac {N_N}{\pi R^2_d} } \int
 d\sigma(\nu_\mu + N \to \mu^- + Q)\; {{d^2 N_\nu} \over { d E_{\nu_\mu}}
 { d E_{\theta}} }\;
 (1-{\cal P}_{\nu_\mu \to \nu_{e,{\tau}}})\; dE_{\nu_\mu}d{\theta}
 \ee
 
 \noindent
 where {\cal P} is the total probability of oscillation of a muon
 neutrino to any other flavour. Here we use the $\nu_\mu
 \longrightarrow \nu_\tau$ oscillation probability corresponding to
 the solution space for the atmospheric $\nu_\mu$ deficit
 \cite{sk_atmos}, and the $\nu_\mu \longrightarrow \nu_e$ oscillation
 probability for the Mikhyev-Smirnov-Wolfenstein (MSW) solution to the
 solar neutrino problem \cite{other_atmos}.  In any case, for a
 near-site detector, the predictions on b-production have no
 perceptible dependence on the precise values of the oscillation
 parameters. In our calculation, we have assumed a target of mass
 $100~T$ containing $N_N~=~6.023 \times 10^{31}$ nucleons.  The
 cross-sectional area of the target ($\pi R_d^2$) and the distance (L)
 from the muon decay point to the target define a cone with the
 semi-vertical angle $\theta_d$, related through the relation $R_d = L
 \theta_d$.  While the available energy range of the the neutrino
 (together with the energy distribution) is specified by the energy of
 the decaying muon, the $\theta$-integration has an upper limit
 $\theta_d$. This brings in not only the detector size but also the
 length of the straight section of the storage ring and the
 probability of the muon decaying at any given point along the
 straight section. This has been taken into account in our
 calculation, where we have assumed a straight section of length
 $100~m$ and that $2 \times 10^{20}$ muons decay within this section
 per year.  Our numerical results correspond to a target cross-section
 of $1~m^2$. Also, CTEQ4LQ \cite{cteq} parton distributions have been
 used here.

%%--------------------------------feynman diagrams
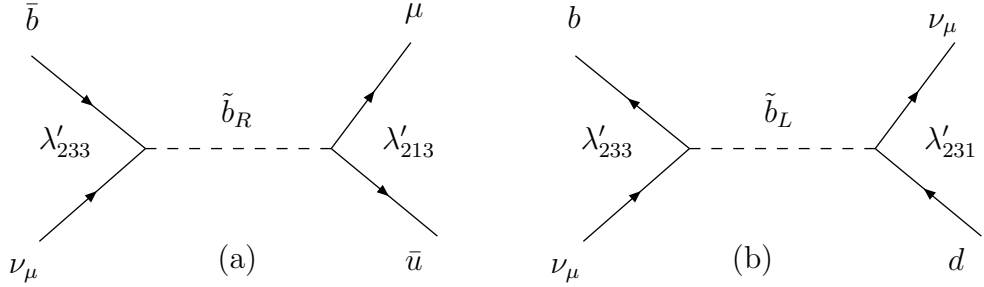
\begin{figure}
\hspace*{-.7in}
%\documentstyle[12pt,psfig,axodraw]{article}
%\begin{document}
%\begin{center}

\begin{picture}(480,100)(0,0)
\vspace*{- 1in}
\ArrowLine(60,-5)(100,30)
\Text(60,-16)[r]{$\nu_\mu$}
\ArrowLine(57,65)(100,30)
\Text(60,80)[r]{${\bar b}$}
\DashLine(100,30)(170,30){4}
\Text(135,40)[b]{${\tilde b}_R$}
\ArrowLine(170,30)(200,70)
\Text(202,80)[b]{${\mu}$}
\ArrowLine(170,30)(210,-3)
\Text(202,-16)[b]{$\bar u$}
\Text(200,30)[b]{$\lambda '_{213}$}
\Text(70,30)[b]{$\lambda '_{233}$}
\Text(135,-15)[b]{(a)}

\ArrowLine(265,-5)(305,30)
\Text(265,-16)[r]{$\nu_\mu$}
\ArrowLine(305,30)(262,65)
\Text(265,80)[r]{$b$}
\DashLine(305,30)(375,30){4}
\Text(340,40)[b]{${\tilde b}_L$}
\ArrowLine(375,30)(405,70)
\Text(402,76)[b]{$\nu_\mu$}
\ArrowLine(415,-3)(375,30)
\Text(407,-16)[b]{$d$}
\Text(405,30)[b]{$\lambda' _{231}$}
\Text(275,30)[b]{$\lambda' _{233}$}
\Text(330,-15)[b]{(b)}

\end{picture}
%\end{center}
%\end{document}
\vskip .3in
\caption{ {\em Feynman diagrams for  processes producing a $b$
 quark or a ${\bar b}$ antiquark in R-parity violating SUSY.}
}
\label{fig:fig1}
\end{figure}
%------------------------------------------------- 

The contributions to the $b$ quark production from the Feynman
diagrams shown in figure 1 are obtained from the following
Fierz-rearranged amplitudes:

\bea
{\cal M}_{R\!\!\!/} (\nu_\mu + {\bar u} \to \mu^- + {\bar b})
& = & {{\lambda'_{213} \lambda'_{233}} \over
 {2 (t - m_{{\tilde b}_R}^2)}}
\big[\bar u_{\mu} \gamma_{\mu} P_L u_{\nu_\mu}\big]\,
\big[\bar v_{u} \gamma^{\mu} P_L v_b \big] \nonumber\\
{\cal M}_{R\!\!\!/} (\nu_\mu + d \to \nu_\mu + b)
 & = & {{\lambda'_{231} \lambda'_{233}} \over
 {2 (t - m_{{\tilde b}_L}^2)}}
\big[\bar u_{\nu_\mu} \gamma_{\mu} P_L u_{\nu_\mu}\big]\,
\big[\bar u_{b} \gamma^{\mu} P_R u_d \big] \nonumber\\ 
\eea
 
%-------------------------------------------------  
\noindent
where, $t = (p^{i} _{\nu_\mu}  - p_b)^2$ in each case.

As can be seen from above, b-production via charged current is driven
by the product $\lambda^{'}_{213} \lambda^{'}_{233}$, while
$\lambda^{'}_{231} \lambda^{'}_{233}$ controls the neutral current
event rate. We take b-squark mass to be $300~GeV$. With
this choice, a not-so-stringent constraint ($\sim 0.048$) exists on
the first one of these products \cite{china_bound}, which is not
appreciably different from the product ($\sim 0.046$) of the upper
limits on the individual couplings taken in isolation \cite{allanach}.
To be conservative, we have used the value corresponding to this
latter limit.  As for the second pair of couplings which controls the
neutral current rate, the product of the individual limits is
approximately $0.14$ for a b-squark of mass $300~GeV$.  A somewhat
weaker limit for the same $m_{\tilde b}$ ($\sim 0.225$) is derived
from the $\nu_\mu\;N \ra \nu_\mu\; b + X$ scattering data from
Fermilab NuTeV Experiment\cite{nutev}.  On the other hand, searches
for the FCNC decay $B^0 \ra \mu^+ \mu^-$ at CDF \cite{cdf} and CLEO
\cite{cleo} set upper limits on the branching ratio of the above decay
at $8.6 \times 10 ^{-7}$ and $6.1 \times 10 ^{-7}$ respectively.  The
BaBar B-factory experiment, too, has a projected upper limit of $5.0
\times 10^{-7}$ that can be obtained on the branching ratio for $B^{0}
\longrightarrow \mu^{+} \mu^{-}$ \cite{babar}. Using the last one of
these, one may be able to set a limit of about $0.018$ on
$\lambda^{'}_{231} \lambda^{'}_{233}$ for the above squark mass.

%---------------------------------------------------------------------
\begin{figure}[ht]
\centerline{ 
\epsfxsize=10cm\epsfysize=7.0cm
                     \epsfbox{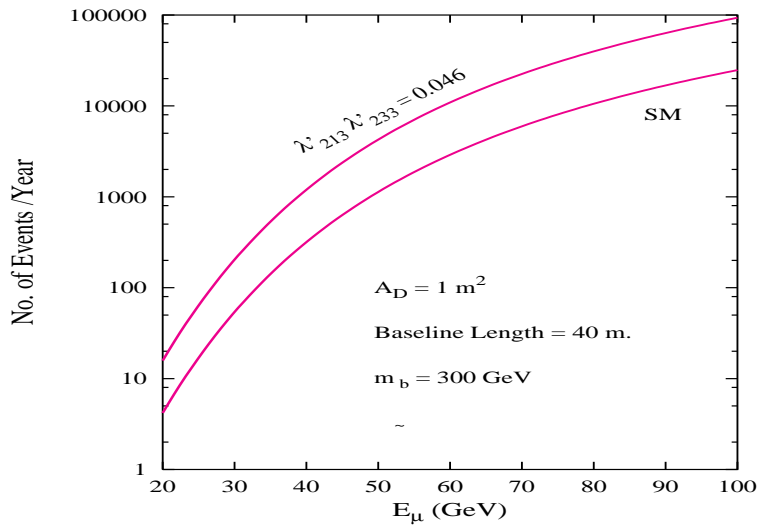}
}

\caption{{\it Number of events coming from 
 $\nu_\mu $-N DIS at a muon storage ring  in charged current interaction
with 100 T of target material.}
}
\label{fig:fig2}
\end{figure}
%-------------------------------------------------------------------

\noindent
The other process which can be of useful in this context is the rare
decay $B \longrightarrow X_d \;\mu^{+} \mu^{-}$.  However, here the
measurement of the decay rate depends crucially on the end-point
analysis of the $\mu^{+} \mu^{-}$ invariant mass spectrum. The
uncertainties coming from the onset of resonant peaks as well as those
in heavy-to-light transition form-factors make the prospect less
bright for strengthening the limit on the new physics effective
interaction to any substantial extent.  Results coming from, say, the
LHC-B experiment (via the channel $B \longrightarrow \rho \;\mu^{+}
\mu^{-}$) may throw some further light on this issue \cite{lhc-b}.
Here we present our numerical results in terms of the current limit on
$\lambda^{'}_{231} \lambda^{'}_{233}$ as well as the projected limit
from B-factories.

In figure \ref{fig:fig2}, we have plotted the number of charged current
b-production events expected per year against the muon energy, for a
$100~T$ target of cross-section $1~m^2$.  The SM (using $V_{ub} =
0.004$ \cite{pdg}) rates as well as those from $R$ parity violating
SUSY have been shown for comparison.  It is clear from the graph that,
with values of the non-standard couplings satisfying the existing
constraints, an excess of 3 to 4 times over the standard model event
rate can be expected, with, say $E_\mu~=~50~GeV$.

%---------------------------------------------------------------------
\begin{figure}[ht]
\centerline{ 
\epsfxsize=10cm\epsfysize=7.0cm
                     \epsfbox{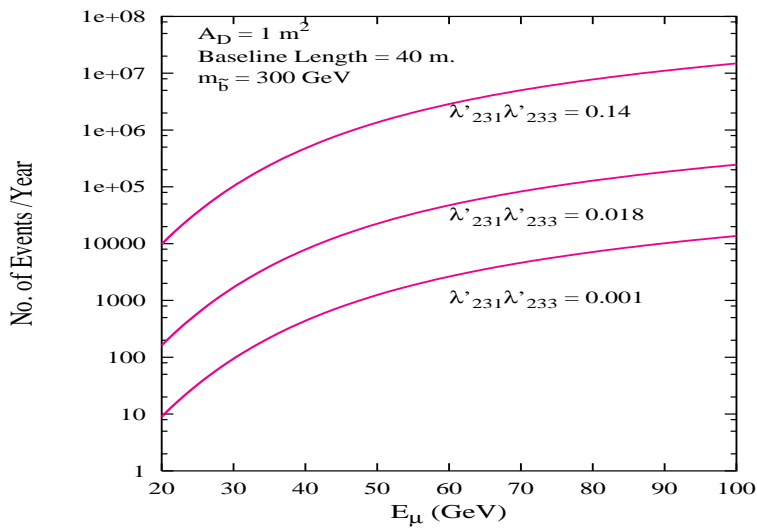}
}

\caption{{\it Number of events coming from 
 $\nu_\mu $-N DIS at a muon storage ring  in neutral current interaction 
 with 100 T of target material.}
}
\label{fig:fig3}
\end{figure}
%-------------------------------------------------------------------

The new physics effects are even more spectacular for neutral current
where the SM contributions are negligibly small. As figure
\ref{fig:fig3} shows, a very large number of neutral current events
with a b in the final state is predicted.  This is true not only with
$\lambda^{'}_{231} \lambda^{'}_{233}$ at the current experimental
upper bound, but also using the limit envisioned at B-factories, and
even with a value one order lower. Even with in the last, rather
conservative, choice, one expects an event rate of about a thousand
per year with $E_{\mu} = 50 ~GeV$, so that a b-tagging efficiency even
as low as 25\% will still make the events detectable.  It can thus be
argued that the observation (or otherwise) of b-production in neutral
current events at a neutrino factory will enable one to probe
flavour-violating coupling of the relevant types (in models including
R-parity violating SUSY) to a higher degree of precision than in the
currently operative B-factories.

For charm production, on the other hand, although the SUSY
contributions can give rise to event rates of observable order of
magnitude with the effective new coupling fixed at the experimental
limit, the SM contributions are larger, since the CKM element $V_{cd}$
is involved in the process. Consequently, It may be somewhat more
difficult to discern the additional contributions. In addition, the
structure of the R-violating couplings implies that at least one
charge $-\frac{1}{3}$ quark superfield must be involved at each
$\lambda^{'}$-type interaction vertex.  Consequently, charm production
in neutral current events is not possible at the tree level.
 
Although we have discussed the above effects in the light of an
R-parity violating SUSY scenario, the general features of our results
are true for any non-standard theory which allows flavour violating
couplings involving a b-quark. It is possible to make predictions for
such models by suitably replacing
${\lambda^{'}\lambda^{'}}/2m_{\tilde{b}}^2$ by the coefficient of the
new effective four-fermion interaction term, on which all the limits
discussed by us will be still applicable.

In summary, we have demonstrated  that  the production of the  b-quark
with substantial  rate at  a neutrino factory   is a  clear signal  of
non-standard interactions  for  neutrinos.   In  particular,   we have
considered  an R-parity violating supersymmetric  theory  to predict 
markedly enhanced b-production rates,    particularly in neutral current
events, with values of the  relevant couplings well below
the existing and projected limits.

\def\baselinestretch{1.8}

%%%%%%%%%%%%%%%%%%%%%%%%%%%%%%%%%%%%%%%%%%%%%%%%%%%%%%%%%%%%%%%%%%%%%%%%
%       %%%     Define Journal macros                                  %
% \newcommand{\araa}[3]{{\em Annu. Rev. Astron. Astrophys.\/}   
%          {\bf#1} (19#3) #2}                                          %
% \newcommand{\ptp}[3]{{\em Prog. Theoret. Phys. (Kyoto)\/} 
%          {\if#1} (19#3) #2}                                          %
\newcommand{\plb}[3]{{Phys. Lett.} {\bf B#1} (#3) #2}                  %
\newcommand{\prl}[3]{Phys. Rev. Lett. {\bf #1} (#3) #2}        %
\newcommand{\rmp}[3]{Rev. Mod.  Phys. {\bf #1} (#3) #2}             %
\newcommand{\prep}[3]{Phys. Rep. {\bf #1} (#3) #2}                     %
\newcommand{\rpp}[3]{Rep. Prog. Phys. {\bf #1} (#3) #2}             %
\newcommand{\prd}[3]{Phys. Rev. {\bf D#1} (#3) #2}                    %
\newcommand{\prc}[3]{{Phys. Rev.}{\bf C#1} (#3) #2}  
\newcommand{\np}[3]{Nucl. Phys. {\bf B#1} (#3) #2}                     %
\newcommand{\npbps}[3]{Nucl. Phys. B (Proc. Suppl.) 
           {\bf #1} (#3) #2}                                           %
\newcommand{\sci}[3]{Science {\bf #1} (#3) #2}                 %
\newcommand{\zp}[3]{Z.~Phys. C{\bf#1} (#3) #2}                 %
\newcommand{\mpla}[3]{Mod. Phys. Lett. {\bf A#1} (#3) #2}             %
 \newcommand{\epj}[3]{{Euor. Phys. J.} {\bf C#1} (#3) #2}          %
 \newcommand{\apj}[3]{ Astrophys. J.\/ {\bf #1} (#3) #2}       %
\newcommand{\astropp}[3]{Astropart. Phys. {\bf #1} (#3) #2}            %
\newcommand{\ib}[3]{{ ibid.\/} {\bf #1} (#3) #2}                    %
\newcommand{\nat}[3]{Nature (London) {\bf #1} (#3) #2}         %
 \newcommand{\app}[3]{{ Acta Phys. Polon.   B\/}{\bf #1} (#3) #2}%
\newcommand{\nuovocim}[3]{Nuovo Cim. {\bf C#1} (#3) #2}         %
\newcommand{\yadfiz}[4]{Yad. Fiz. {\bf #1} (#3) #2;             %
Sov. J. Nucl.  Phys. {\bf #1} #3 (#4)]}               %
\newcommand{\jetp}[6]{{Zh. Eksp. Teor. Fiz.\/} {\bf #1} (#3) #2;
           {JETP } {\bf #4} (#6) #5}%
\newcommand{\philt}[3]{Phil. Trans. Roy. Soc. London A {\bf #1} #2  
        (#3)}                                                          %
\newcommand{\hepph}[1]{ hep--ph/#1}           %
\newcommand{\hepex}[1]{     hep--ex/#1}           %
\newcommand{\astro}[1]{     astro--ph/#1}         %
%       \relax                                                         %
%       %%%     End     Journal macro definitions                      %
%                                            %x
%%%%%%%%%%%%%%%%%%%%%%%%%%%%%%%%%%%%%%%%%%%%%%%%%%%%%%%%%%%%%%%%%%%%%%%%

\end{document}